\documentclass[12pt,preprint]{aastex}

\begin{document}
\title{The LEAP of Pulsars in the Milky Way}

\author{M. M. McKinnon}
\affil{National Radio Astronomy Observatory\altaffilmark{1}\altaffiltext{1}
{The National Radio Astronomy Observatory is a facility of the National 
Science Foundation operated under cooperative agreement by Associated 
Universities, Inc.} Socorro, NM, USA}

\begin{abstract}
The location of objects on the celestial sphere is a fundamental 
measurement in astronomy, and the distribution of these objects within 
the Milky Way is important for understanding their evolution as well 
as the large scale structure of the Galaxy. Here, physical concepts 
in Galactic astronomy are illustrated using straightforward mathematics 
and simplifying assumptions regarding the geometry of the Galaxy. 
Specifically, an analytical model for a smooth distribution of 
particles in an oblate ellipsoid is used to replicate the observed 
distributions of the Galactic coordinates for pulsars and supernova 
remnants. The distributions and the Lambert equal area projections 
(LEAPs) of the coordinates suggest that the dominant factors determining 
the general shape of the distributions are the heavy concentration of 
objects in the Galactic plane and the offset of the Galactic center 
from the coordinate system origin. The LEAPs and the distributions also 
show that the dispersion of pulsars about and along the plane are much 
larger than that for their progenitor supernovae. Additionally, the model 
can be used to derive an analytical expression for the dispersion measure 
along any line of sight within the Galaxy. The expression is used to create 
a hypothetical dispersion measure-distance map for pulsars in the Galaxy.

\end{abstract}


\section{Introduction}

The distribution of objects in the Milky Way is important for understanding 
their formation and evolution, as well as the large scale structure of the 
Galaxy. Although Galactic coordinates are measured with respect to the 
location of the Sun, most analytical models for the locations of objects in 
the Galaxy use a coordinate system having an origin at the Galactic center 
(GC; Sartore et al. 2009; Lorimer et al. 2006). And although the locations 
are measured in a spherical coordinate system, the models are formed using 
cylindrical coordinates, with the height, azimuth, and equatorial radial 
distance of the objects taken as independent random variables (RVs). The 
distribution for azimuth is always uniform over $2\pi$, due to the symmetry 
provided by a galactocentric origin. The height of an object above the 
Galactic plane (GP) is almost always treated as an exponential RV (Lorimer 
et al. 2006; Bahcall 1986; Lyne et al. 1985; Gunn \& Ostriker 1970). Most 
models assume the equatorial radii of stellar coordinates follow gamma or 
exponential distributions (Lorimer et al. 2006; Bahcall 1986), while others 
invoke a normal distribution (Sartore et al. 2009). The parameterization of 
these models requires reasonably accurate estimates of the distances to the 
objects for a proper transformation of coordinates between the solarcentric 
and galactocentric coordinate systems. The distance estimates for pulsars 
(PSRs) are generally made from a model of the distribution of free electrons 
within the Galaxy (Taylor \& Cordes 1993; Cordes \& Lazio 2008) in 
combination with the observed value of a PSR's dispersion measure. The 
distance estimates can introduce a significant source of error in the model 
parameterization because actual measurements of PSR distances via parallax 
have shown some estimates to be in error by a factor of two or more 
(Cordes \& Lazio 2008; Brisken et al. 2002). The objective of this paper 
is to develop a solarcentric analytical model of spheroidal Galactic 
coordinates that can be parameterized without direct knowledge of the 
distances to objects in the Galaxy.

\begin{figure}
\plotone{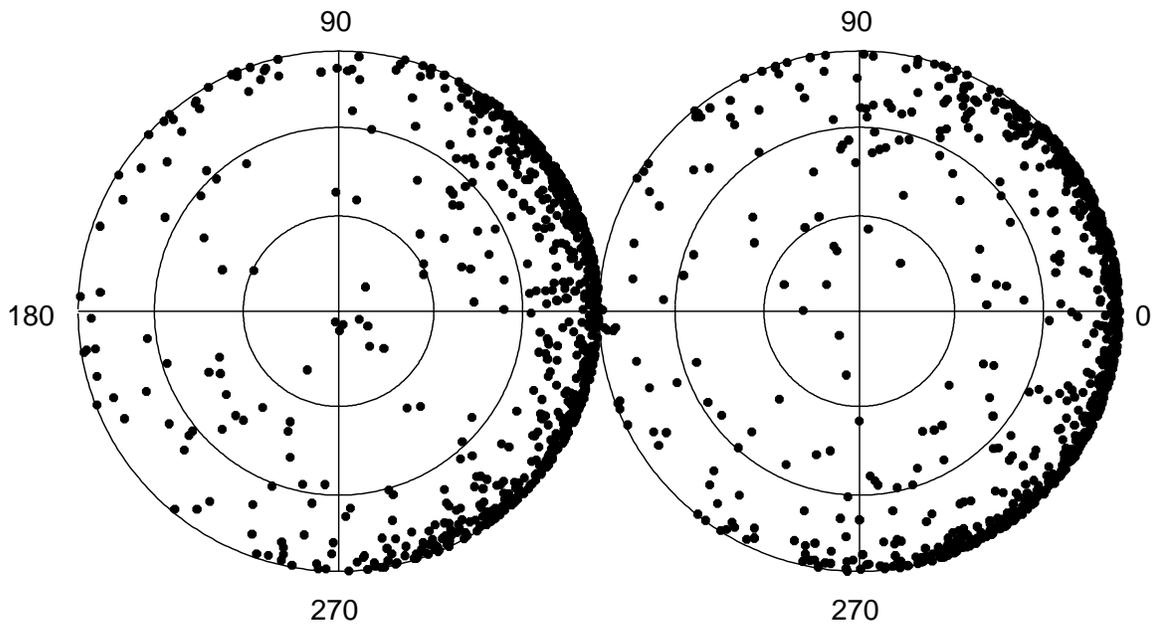}
\caption{Lambert equal-area projection of pulsar Galactic coordinates.
The perimeter of the figure corresponds to the Galactic plane. The 
concentric circles are lines of constant Galactic colatitude.} 
\label{fig:psrleap}
\end{figure}

\begin{figure}
\plotone{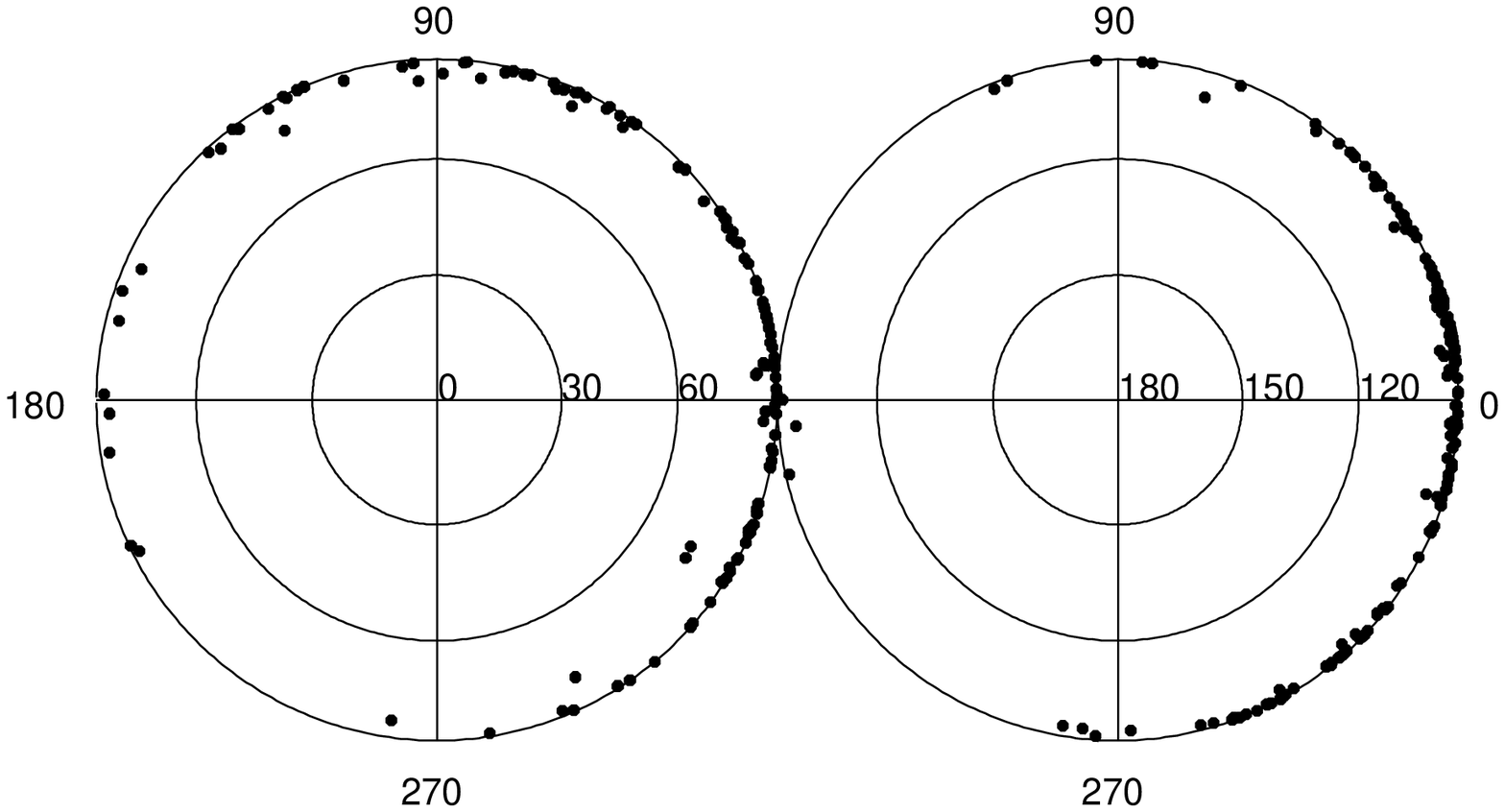}
\caption{Lambert equal-area projection of supernova Galactic coordinates. 
The numbers on the perimeter of the figure denote Galactic longitude. 
The numbers labelling the interior concentric circles denote Galactic 
colatitude.}
\label{fig:snrleap}
\end{figure}

The distribution of objects within the Galaxy is perhaps best illustrated 
by a Lambert equal area projection (LEAP), which is a polar plot of the 
Galactic coordinates of the objects in the two hemispheres of the 
celestial sphere (e.g. see Figs.~\ref{fig:psrleap} and~\ref{fig:snrleap}). 
The main advantage of a LEAP is it preserves the density of data points 
in the projection, unlike other projection methods (e.g. orthographic) 
which do not (Fisher et al. 1987). A LEAP consists of two sets of concentric 
circles centered on the location of the Sun. The left set of circles in 
a LEAP is the projection as viewed from above the GP, and the right set 
is the view from below the GP. The azimuth and radius of the points in 
the polar plots represent an object's Galactic coordinates. Galactic 
longitude, $\phi$, is equal to a point's azimuth in a LEAP and increases 
in the counterclockwise direction from the GC, which is located at the 
far right of each circle set. A point's radius in the polar plot is equal 
to $2\sin(\theta/2)$, where $\theta$ is the Galactic colatitude of the 
object. Objects located in the GP fall on the perimeter of the LEAP, and 
those located off the GP reside within the perimeter.

The LEAPs of PSRs and supernova remnants (SNRs) are shown in 
Figures~\ref{fig:psrleap} and Figure~\ref{fig:snrleap}, respectively. Since 
PSRs are born in supernovae, one might expect their distributions on the 
sky to be similar. Both objects reside primarily in the GP in what is 
known as a girdle-type distribution in the statistical literature (Fisher 
et al. 1987). The LEAPs show the dispersion of PSRs perpendicular to the 
GP is much greater than that of their progenitor SNRs. The same may also 
be true of the dispersion along the GP. 

\section{Statistical Model of Galactic Coordinates}

An analytical, statistical model was developed to derive a joint 
probability density for the Galactic coordinates of any type of object 
(e.g. PSR, SNR, star, HII region, globular cluster, etc.) in the Galaxy. 
The motivation for developing the model is to quantify and explore the 
differences in the spatial distributions of PSRs and SNRs. 

The model's coordinate system has an origin corresponding to the location 
of the Sun. Its x-axis lies in the direction of the GC, and its z-axis 
is perpendicular to the GP. The model assumes the Cartesian coordinates 
of the objects are independent, normal RVs. The assumption of normal RVs 
allows an analytical solution to be derived because the coordinate 
distributions are continuous at all locations within the model galaxy. 
The RV representing the z-coordinate has a zero mean and a standard 
deviation $\sigma_z=\sigma$, which is the dispersion of the objects 
{\it about} the GP. The RV representing the x-coordinate has a mean equal 
to the distance, $R_o$, between the Sun and the GC and a standard deviation
of $\sigma_x=\sigma(1+\rho^2)^{1/2}$, which is the dispersion of the 
objects {\it along} the GP. The RV for the y-coordinate has a zero mean 
and a standard deviation equal to $\sigma_x$. By construction, the model 
inherently assumes the Sun resides precisely in the GP. As with most other 
analytical models of the Galaxy, the model neglects the structure of the 
Galaxy's spiral arms. 

The model imposes the form of an oblate ellipsoid on the three-dimensional 
distribution of objects in the Galaxy. The ellipsoid is centered on the GC
and has an axial ratio equal to $(1+\rho^2)^{1/2}$. The model produces a 
smooth, yet continuously varying, spatial density of objects in the Galaxy. 
An analytical expression for the density, $n$, as a function of distance, 
$r$, longitude, and colatitude can be computed from the model assumptions 
by multiplying the probability densities for the x, y, and z coordinates by 
one another, since they are independent RVs, and converting these Cartesian 
coordinates to spherical ones. The density is

\begin{eqnarray}
n(r,\theta,\phi) & = & {1\over{\sigma^3(1+\rho^2)(2\pi)^{3/2}}}
    \exp{\Biggl[{-R_o^2(1+\rho^2\cos^2\theta-\sin^2\theta\cos^2\phi)
    \over{2\sigma^2(1+\rho^2)(1+\rho^2\cos^2\theta)}}\Biggr]} \nonumber \\
& \times & \exp{\Biggl\{{-(1+\rho^2\cos^2\theta)\over{2\sigma^2(1+\rho^2)}}
  \Biggl[r-{R_o\sin\theta\cos\phi\over{(1+\rho^2\cos^2\theta)}}\Biggl]^2
  \Biggr\}}.
\end{eqnarray}
As written, the equation for density is normalized such that it integrates 
to unity over volume. The actual number density is the density multiplied 
by a constant equal to the total number of objects in the Galaxy.

\subsection{Joint Probability Density of Galactic Coordinates}

The joint probability density of an object's Galactic colatitude and 
longitude can be calculated by integrating the density over radius. 
The joint probability density is

\begin{eqnarray}
f(\theta,\phi) & = & \sin\theta\int_0^\infty n(r,\theta,\phi)r^2dr 
  \nonumber \\
 & = & u\exp\Biggl[-{s^2\over{2(1+\rho^2)}}\Biggr]
 \Biggl\{\exp{\Biggl({v^2\over{2}}\Biggr)} \Biggl[1+{\rm erf}
 \Biggl({v\over{\sqrt{2}}}\Biggr)\Biggr](1 + v^2)+v\sqrt{{2\over{\pi}}}
 \Biggr\},
\end{eqnarray}
where $\rm{erf}(x)$ is the error function, and $u$ and $v$ are functions of 
$\theta$ and $\phi$ given by
\begin{equation}
u(\theta) = {\sin\theta\over{4\pi}}
                 {(1+\rho^2)^{1/2}\over{(1 +\rho^2\cos^2\theta)^{3/2}}}
\end{equation}
\begin{equation}
v(\theta,\phi) = {s\sin\theta\cos\phi\over{
                 [(1+\rho^2)(1 +\rho^2\cos^2\theta)]^{1/2}}}
\end{equation}
The joint density is a function of two free parameters, $s=R_o/\sigma$ 
and $\rho$. The inverse of the parameter $s$, in units of radians, is 
the angular dispersion of the objects in Galactic colatitude. The 
quantity $(1+\rho^2)^{1/2}/s$, again in units of radians, is related to 
the angular dispersion in Galactic longitude. The colatitude distribution 
narrows with increasing $s$. The longitude distribution narrows with 
increasing $s$ and decreasing $\rho$. A fit of the distribution of an 
object's coordinates to Equation 2 does not require distance information 
because distance was eliminated in the calculation of the angles' joint 
probability density with the integration over $r$.

\subsection{Pulsar Dispersion Measure}

The dispersion measure (DM) of a PSR is the density of free electrons in 
the interstellar medium integrated over the distance, $d$, to the PSR. The 
model developed here specifies the position-dependent density of any 
object in the Galaxy. One can compute an analytical expression for DM 
along any line of sight by assuming the objects in the model are free 
electrons. The DM is 

\begin{eqnarray}
{\rm DM}(d,\theta,\phi) & = & N_e\int_0^dn(r,\theta,\phi)dr \nonumber \\
  & = & {N_e\over{4\pi\sigma^2}}{\sqrt{2}\over{g(\theta)}}
  \exp\Biggl[-{R_o^2(1+\rho^2\cos^2\theta-\sin^2\theta\cos^2\phi)\over
  {\sigma^2g^2(\theta)}}\Biggr] \nonumber \\ & \times &
  \Biggl\{{\rm erf}\Biggl[{d(1+\rho^2\cos^2\theta)-R_o\sin\theta\cos\phi)
  \over{\sigma g(\theta)}}\Biggr] + {\rm erf}\Biggl[{R_o\sin\theta\cos\phi
  \over{\sigma g(\theta)}}\Biggl]\Biggl\},
\end{eqnarray}
where $N_e$ is the total number of free electrons in the Galaxy, and 
$g(\theta)$ is a function of Galactic colatitude given by
\begin{equation}
g(\theta) = [2(1+\rho^2)(1+\rho^2\cos^2\theta)]^{1/2}.
\end{equation}

\section{Data Analysis}

Two-dimensional histograms of the colatitude and longitude for PSRs and 
SNRs were computed using coordinates given in the ATNF PSR catalog 
(Manchester et al. 2005) and Green's (2004) catalog of Galactic SNRs. 
The PSR data were edited to exclude extragalactic PSRs and multiple PSRs 
in globular clusters. Estimates of $s$ and $\rho$ were made with a 
two-dimensional, least squares fit of the histograms to the joint 
probability density given by Equation 2. 

\begin{figure}
  \plotone{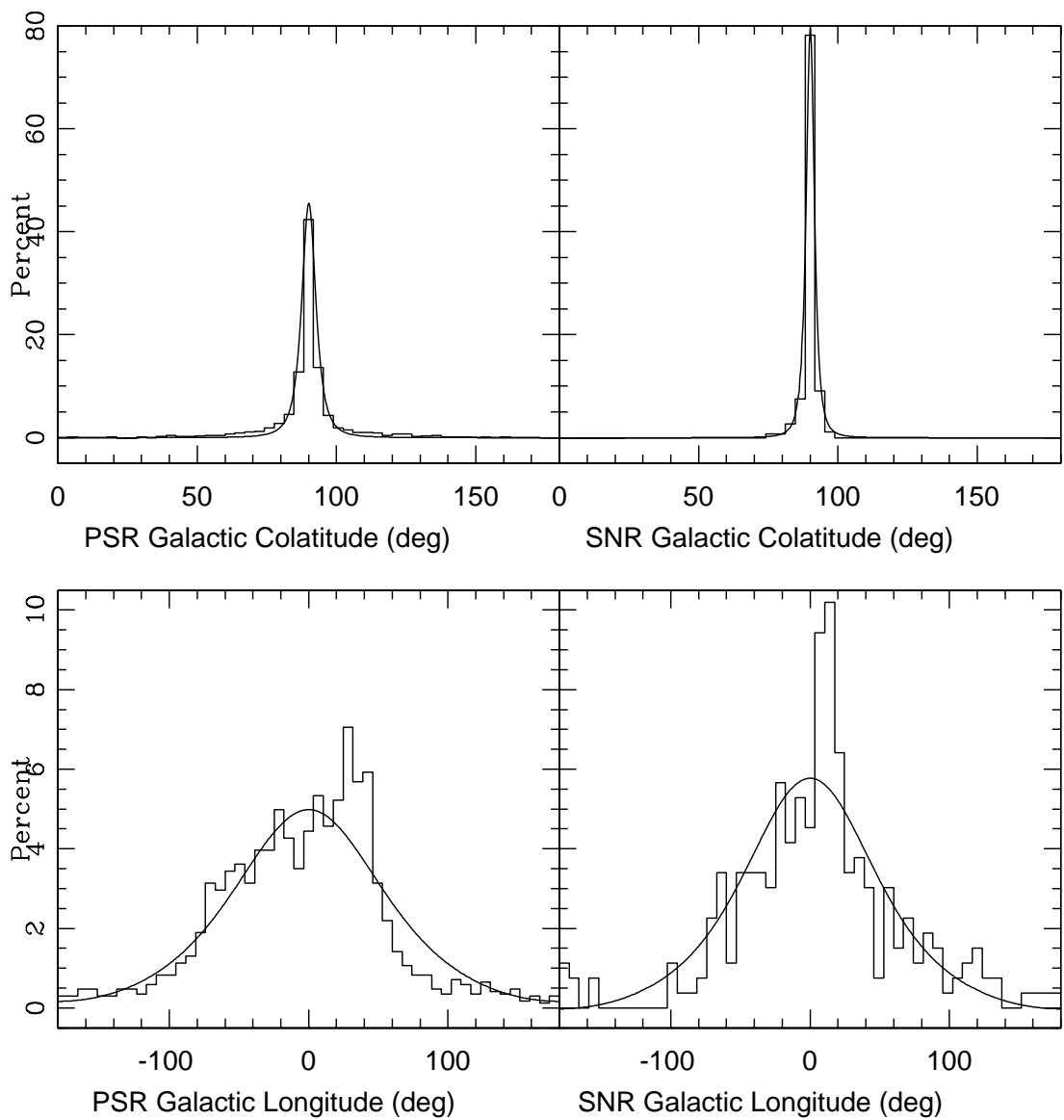}
  \caption{Histograms of the Galactic colatitude (top panel) and 
   longitude (bottom panel) for pulsars and supernova remnants. The 
   histograms were computed from the measured coordinates of the objects. 
   The continuous solid line in each panel denotes the best fit of 
   Equation 2 to the two-dimensional histogram of each object's 
   coordinates.}
  \label{fig:coordhist}
\end{figure}

The results of the fit are shown in Table 1 and Figure~\ref{fig:coordhist}. 
The entries in the table are the number of objects in each fit, N, the fit 
parameters, $s$ and $\rho$, the axial ratio of each object's oblate 
ellipsoid, $(1+\rho^2)^{1/2}$, and the dispersions perpendicular and 
parallel to the GP as calculated from $s$ and $\rho$ assuming $R_o=8.5$ 
kpc. Figure~\ref{fig:coordhist} shows histograms of colatitude and 
longitude computed from the PSR and SNR catalog data. Theoretical 
distributions of $\theta$ and $\phi$ were calculated by numerically 
integrating the joint density over $\phi$ and $\theta$, respectively, 
using the values of $s$ and $\rho$ obtained from the least squares fits. 
The continuous solid line in each panel of the figure shows the fitted 
theoretical distribution.

To explore the verstatility of the model, the same analysis was applied
to the Galactic coordinates for globular clusters using data from the
catalog maintained by Harris (1996). The results of the analysis
are summarized in the last row of Table 1. The LEAP and coordinate 
histograms for globular clusters are not shown.

\begin{deluxetable}{lrccccc}
\tablewidth{400pt}
\tablecaption{Derived Parameters for the Joint Probability Density of Galactic 
Coordinates for Pulsars, Supernova Remnants, and Globular Clusters}
\tablehead{\colhead{Object} & \colhead{N} & \colhead{$s$} & \colhead{$\rho$} & 
\colhead{$(1+\rho^2)^{1/2}$} & \colhead{$\sigma_z$ (kpc)} & 
\colhead{$\sigma_x$ (kpc)}}
\startdata
  Pulsars & 1687 & 10.4 & 12.7 & 12.8 & 0.8 & 10.4 \\
  Supernova Remnants &  265 & 21.8 & 19.2 & 19.2 & 0.4 &  7.5 \\
  Globular Clusters & 150 & 3.8 & 0.6 & 1.2 & 2.2 & 2.6 \\
\enddata
\end{deluxetable} 

\begin{figure}
  \plotone{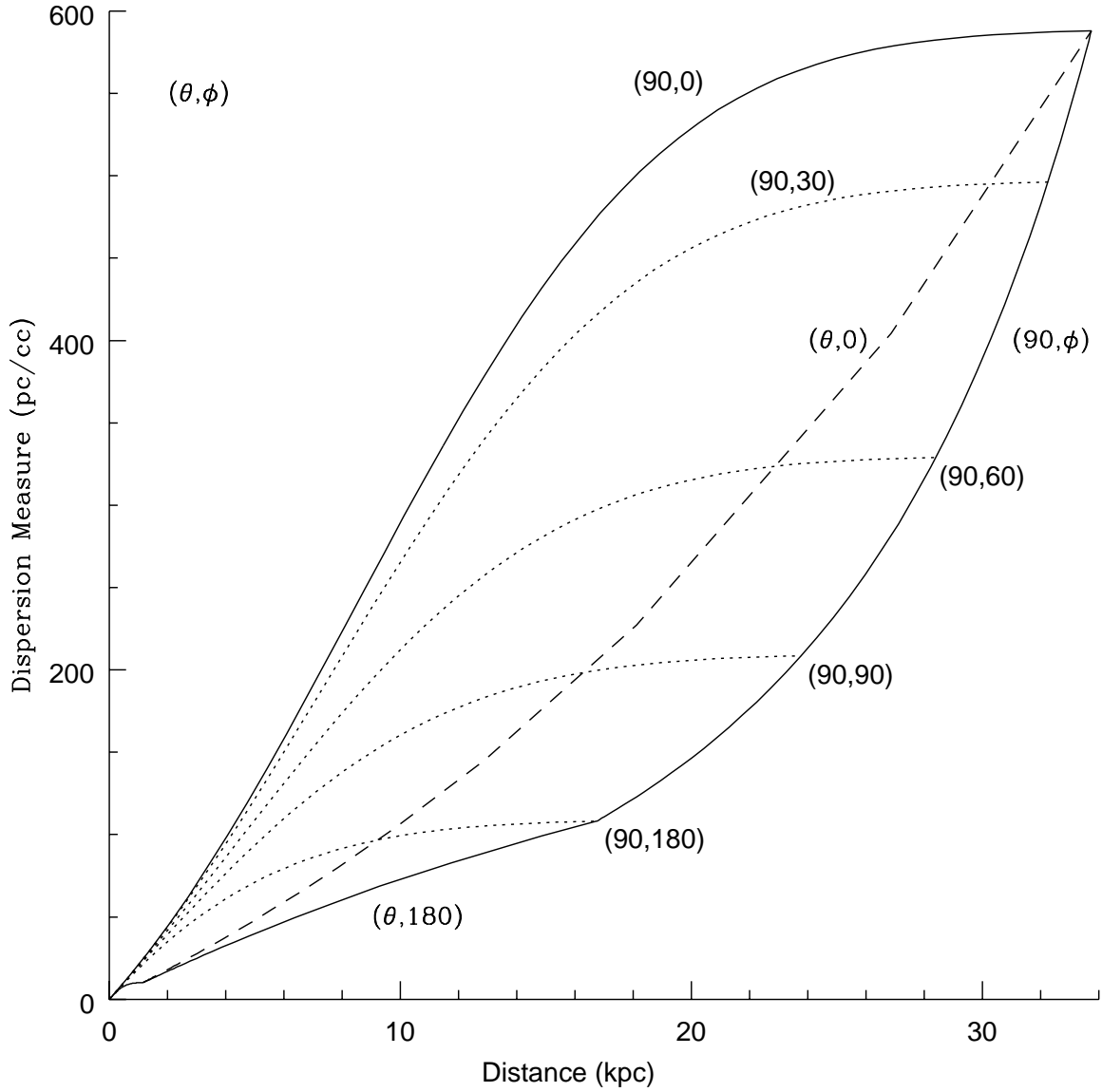}
  \caption{Hypothetical dispersion measure-distance map for pulsars in 
   the Galaxy. The numbers in parenthesis denote the colatitude, $\theta$,
   and longitude, $\phi$, in units of degrees, for different lines
   of sight through the Galaxy.}
  \label{fig:map}
\end{figure}

Equation 5 was used to develop a hypothetical DM-distance map for PSRs in 
the Galaxy (Fig.~\ref{fig:map}). The rather arbitrary parameters used to 
produce the map were $R_o=8.5$ kpc, $\sigma=0.4$ kpc, $\rho=20$, and 
$N_e/\sigma^2=9\times 10^4\ {\rm pc/cm^3}$. A physical boundary for the 
Galaxy had to be estimated to produce the map. The boundary is the edge 
of the Galaxy, or maximum distance, $r_{max}$, along any line of sight. 
It was estimated from the equation for the surface of an ellipsoid with 
the maximum extent of the Galaxy taken to be $3\sigma$ from the GC in 
the z-direction and $3\sigma(1+\rho^2)^{1/2}$ from the GC in the x- and 
y-directions. The maximum distance is 
\begin{equation}
r_{max}(\theta,\phi) = {R_o\sin\theta\cos\phi + 
  \{R_o^2\sin^2\theta\cos^2\phi + 
  [9\sigma^2 (1+\rho^2)-R_o^2][\sin^2\theta+(1+\rho^2)\cos^2\theta]\}^{1/2}
  \over{\sin^2\theta + (1+\rho^2)\cos^2\theta}}.
\end{equation}
The solid lines in the figure show the Galactic boundary in DM-distance 
space. Ideally, all values of DM and distance for PSRs in the Galaxy would 
reside within this boundary. The upper boundary gives the maximum DM as a 
function of distance for PSRs residing in the direction of the GC. It 
terminates at the far side of the Galaxy on the right side of the figure. 
The lower right boundary of the map is the DM measured at the edge of the 
Galaxy for lines of sight within the GP having different values of 
longitude. (When $\theta=\pi/2$, the lower boundary in the map is the same 
for longitudes in the range $0\le\phi\le\pi$ as it is for the longitudes 
$2\pi\ge\phi\ge\pi$ due to the geometric symmetry of the problem). The 
cusp in the boundary at a distance of about 17 kpc occurs at the Galactic 
anti-center. The lower left boundary in the map is the DM-distance 
relation for lines of sight with longitude fixed at $\phi=\pi$ and 
colatitude varying in the range $0\le\theta\le\pi/2$. The cusp near 
$d\simeq 1$ kpc is the maximum DM at the edge of the Galaxy directly
above and below the Sun. The small boundary spur in the range $0<d\le 1$ 
kpc is the DM measured between the Sun and the edge of the Galaxy
above and below it. The dotted lines in the map show how the DM varies 
with distance for lines of sight within the GP having fixed values of 
longitude. The dashed line is the maximum DM as a function of distance 
when the longitude of the line of sight is fixed at $\phi=0$ with the 
colatitude varying between $0$ and $\pi/2$.

\section{Results and Discussion}

The angular dispersion of colatitude can be calculated from the inverse of 
the value of $s$ obtained in the model fit. From the entries in Table 1, 
the calculation gives an angular dispersion in colatitude of about 5.5 
degrees for PSRs and about 2.6 degrees for SNRs. The angular dispersion in 
longitude can be calculated from the fit values of $s$ and $\rho$. The 
dispersion in longitude for PSRs and SNRs is 71 degrees and 50 degrees, 
respectively. Summarizing the fit results for PSRs by stating them another 
way, most observed PSRs reside within Galactic longitudes of $\pm 71$ 
degrees and within Galactic latitudes of $\pm 5.5$ degrees.

\begin{figure}
  \plotone{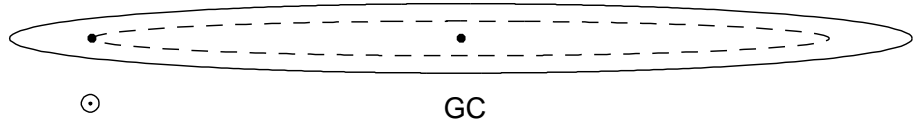}
  \caption{$1-\sigma$ contours for the distribution of pulsars (solid line)
           and supernova remnants (dashed line) within the Galaxy. The
           contours are drawn to scale.}
  \label{fig:contour}
\end{figure}

The entries in Table 1 indicate that the three dimensional spatial 
distributions of PSRs and SNRs are consistent with oblate ellipsoids. The 
PSR ellipsoid has an axial ratio of about 13, while the SNR axial ratio is 
about 19. A representation of the \lq\lq one-sigma" contours of the 
distributions of PSRs and SNRs in the Galaxy is drawn to scale in 
Figure~\ref{fig:contour}. The two dots represent the locations of the GC 
and the Sun. Their separation of $R_o=8.5$ kpc sets the scale for the 
figure. The figure shows that the Sun resides on the edge of the SNR 
contour. 

The entries in Table 1 show the dispersion of PSRs perpendicular to the 
GP is about a factor of two larger than that for SNRs. The difference is 
likely caused by the kick velocity acquired by the PSR in the explosion 
of its progenitor supernova (Gunn \& Ostriker 1970). The kick accounts for 
the high proper motions of PSRs and their locations outside the GP. The 
dispersion of PSRs along the GP is about 40\% larger than that for SNRs. 
It may be caused by a component of PSR kick velocity along the GP. 

The analysis presented here likely underestimates the actual value of $s$ 
and overestimates the actual value of $\rho$ for PSRs. Searches for PSRs 
rely on the impulsive nature of PSRs for their detection. Multipath 
scattering in the interstellar medium can significantly broaden the 
pulse, making them more difficult to find, particularly towards the GC 
where the scattering can be severe. Consequently, PSRs are undersampled 
towards the GC and the actual distribution of PSRs is more heavily 
populated at $\theta=\pi/2$ and $\phi=0$ than what is shown in 
Figure~\ref{fig:coordhist}. This selection effect causes the fitted value 
of $s$ to be smaller than its actual value and the fitted value of $\rho$ 
to be larger. Thus, the calculated value of $\sigma_z$ should be regarded 
as an upper limit since it varies inversely with $s$ and the actual value 
of $s$ is likely larger than the fitted one. By a similar argument, the 
fitted value of $\sigma_x$ should also be regarded as an upper limit.

The results obtained by fitting observed coordinates to the model are 
reasonably consistent with those produced from other models of the Galaxy. 
The major components comprising the large scale structure of the Galaxy 
include, at a minimum, a spheroid of old stars and globular clusters and 
a thin disk of younger stars and atomic and molecular gas 
(Bahcall 1986; Cox 2000). The Galaxy may contain other components, such 
as a \lq\lq thick" disk, a central bulge, and a halo of unseen stars. For the 
spheroid, Bahcall (1986) lists an effective radius of 2.7 kpc and 
a major-to-minor axial ratio of about 1.25. When applied to globular 
clusters, the model described in this paper gives a spheroid effective 
radius in the range of 2.2 to 2.6 kpc and a spheroid axial ratio of 1.2. 
For the disk, Bahcall gives an exponential scale height of 0.3 kpc. Here, 
the dispersion in SNRs about the GP is estimated at 0.4 kpc. The estimate 
of 0.8 kpc for the dispersion of PSRs about the GP may be more consistent 
with the scale height (1.3 kpc) Bahcall lists for a thick disk.

In principle, those PSRs having accurate DM and distance measurements 
(Brisken et al. 2002; Chatterjee et al. 2009) could be used to calibrate 
the model for the distribution of free electrons in the Galaxy by a 
non-linear least squares fit of the data to Equation 5. The fit was made, 
but the resulting values of $\chi^2$ did not decrease as more measurements 
were included in the fit. This means Equation 1 is not the most accurate 
representation of the electron density, which is not surprising since 
we know the distribution of electrons in the Galaxy is not smooth and 
is heavily influenced by the presence of the spiral arms (Cordes \& Lazio 
2008). Nonetheless, the DM relationship (Eqn. 5) is useful for a conceptual 
illustration of how DM might vary with different views through the 
Galaxy (e.g. Fig.~\ref{fig:map}).

The model presented here could be extended to other astrophysical 
problems, such as the emission measure (EM) along different lines of
sight within the Galaxy, DM variations in globular clusters, and column 
density measurements of proto-stellar cores. The EM can be easily 
calculated by squaring the electron density given by Equation 5 and 
integrating the result over distance. Multiple PSRs have been discovered 
in globular clusters (Ransom et al. 2005; Freire et al. 2003), and it may 
be possible to probe the electron density profile of a cluster with this 
model if the differences in the DMs and angular separations of the PSRs in 
the cluster are measureable. Only minor revisions to the model would be 
required, such as compensating the model coordinate system for the location 
of the globular cluster and approximating the cluster geometry as a sphere 
(i.e. by setting $\rho=0$). The column density of a proto-stellar core 
is the density integrated over a specific line of sight through the 
entire core (Dapp et al. 2009). The calculation of column density is very 
similar to the calculation of maximum DM presented in this analysis. As 
with the globular cluster analysis, the model would need to be revised 
slightly to accommodate the viewing geometry of the core. One could 
incorporate different density profiles in the model and evaluate them in a 
comparison with the column density measurements.

\section{Conclusions}

An analytical model with a solarcentric, spherical coordinate system 
was developed for the distribution of objects in the Galaxy in an 
attempt to circumvent complications introduced by the galactocentric, 
cylindrical coordinate system used in other models. Unlike these 
previous models, the present model does not require distance estimates 
for its parameterization. The model reasonably describes histograms of 
measured coordinates for PSRs and SNRs and was used to quantify the 
differences in their distribution parameters. The dispersion of PSRs 
perpendicular to and along the GP is larger than that for SNRs. The 
difference is likely caused by the kick velocity acquired by a PSR in 
the explosion of its progenitor SNR. SNRs reside in a thin disk of the 
Galaxy, while PSRs are located in a thicker disk. When applied to 
globular clusters, the model shows the clusters reside in a spheroid 
centered on the GC. The model can be applied to other objects to 
quantify their spatial distribution and perhaps the large scale 
structure of the Galaxy.

\end{document}